# Microstructure evolution and characteristics of laser-clad lightweight refractory $Nb_xMo_{0.5}Ti_{1.5}Ta_{0.2}Cr$ high-entropy alloy

C.Y. Cui, H.H. Xu, J. Yang, X.G. Cui[*]

School of Mechanical Engineering, Jiangsu University, Zhenjiang 212013, PR China

**Abstract**: Lightweight refractory high-entropy alloy coatings (RHEAcs) of $Nb_xMo_{0.5}Ti_{1.5}Ta_{0.2}Cr$ (where x=1, 1.3, 1.5, and 2) were fabricated on the surface of 316L stainless steel using laser cladding (LC) technology. A comprehensive study was conducted to elucidate the effect of Nb content on the microstructure, microhardness and wear resistance of $Nb_xMo_{0.5}Ti_{1.5}Ta_{0.2}Cr$ RHEAcs before and after annealing at 900°C for 10 h. The results show that the grains are gradually refined with the increase of Nb content. The coating consists mainly of a body-centered cubic (BCC) solid solution, C15-Laves phase, and a small amount of hexagonal close-packed (HCP) solid solution containing Ti. The microhardness of RHEAcs is significantly higher compared to the base material. Notably, at $Nb_{1.3}$, due to the influence of lattice dislocations, the microhardness reaches a peak of 1066.5 HV, which is about 7.11 times higher than that of the base material. On the contrary, at $Nb_2$, the microhardness is at its lowest point, averaging 709.31 HV, but 4.72 times that of the base material. After annealing, an increase in microhardness is observed at all Nb concentrations, up to 31.2% at $Nb_2$. Before annealing, the wear resistance of RHEAcs was slightly better than that of 316L stainless steel at different Nb contents. However, after annealing, the coefficient of friction (COF) and wear rate of the coatings are lower than those of annealed 316L stainless steel, highlighting their excellent wear resistance. It is noteworthy that the loss of wear properties after annealing at Nb1 is at a minimum, obtaining the most balanced wear resistance before and after annealing. The enhanced wear resistance after annealing can be attributed to the presence of ultra-fine grain oxide friction layers, mainly composed of $TiO_2$ and $Ta_2O_5$. The main mode of wear is oxidative wear, with a small amount of wear from abrasive wear.

---

[*]Corresponding author Tel.: +86-511-88797898, Fax: +86-511-88780241.
 E-mail: ccyyy981206@163.com

# 1. Introduction

316L stainless steel is known for its excellent high-temperature properties and remarkable plasticity for easy hot and cold machining [1]. However, its low wear resistance and hardness at elevated temperatures limit its use in high-temperature components [2].

Laser cladding technology has become the preferred method to enhance the surface properties of 316L. Compared with traditional thermal spray coatings and physical vapor deposition techniques, laser cladding technology has the advantage of rapid cooling, which usually results in ultra-fine grain structure and unique phase formation, thus significantly enhancing its mechanical properties, high-temperature resistance and corrosion resistance [3-5]. Therefore, there is a great potential for the application of high hardness, high wear and high temperature resistant coatings on 316L stainless steel components [6-7].

Yeh et al [8] introduced the concept of high entropy alloys (HEAs), which exhibit excellent properties upon formation, such as high hardness, high strength, excellent resistance to softening at high temperatures, and commendable wear and corrosion resistance [9-12]. Refractory high-entropy alloys (RHEAs), an important subset of HEAs, have recently attracted considerable research interest.Senkov et al. proposed TiZrHfNbTa RHEAs [13-19], which were promoted to be short-range chemically ordered by the addition of Hf and Zr, resulting in a reduced density of 9.94 g/cm³ and a room-temperature ductility of more than 50%. However, its high-temperature yield strength is only 92 MPa at 1200°C. Juan et al [20] proposed

TiZrHfNbTaMo, which balances the high-temperature yield strength (556 MPa at 1200°C) and room-temperature ductility ($\varepsilon_f$ = 12%), but still faces the problem of a high density of 9.97 g/cm³. Wen et al. utilized the transformation-induced plasticity (TRIP) [21] concept and changed the molar contents of Nb and Ta in HfNb$_x$Ta$_{0.2}$TiZr [22] to achieve a phase transition from stabilized BCC to BCC + HCP, which resulted in excellent strength (983 MPa) and uniform ductility (>26%). At a Nb content of 0.25 (at%), the theoretical density drops to 8.64 g/cm³. However, its oxidation and wear resistance at high temperatures are poor. Therefore, controlling the friction and oxidation products of the alloy at elevated temperatures by compositional design is the key to improving its high-temperature wear resistance.

In this study, the elemental composition and molar ratios in RHEAS were optimized using first-principles calculations, and a series of coatings of Nb$_x$Mo$_{0.5}$Ti$_{1.5}$Ta$_{0.2}$Cr (with x-values of 0.5, 1, 1.3, 1.5, and 2) lightweight RHEAs were designed and fabricated using laser cladding. In view of the significant thermal efficiency ratio ($T_m/\rho$) and intrinsic toughness of Nb [24], the Nb content was experimentally tuned with the aim of improving the density, strength and high temperature properties of the alloy. Comparative hardness, wear resistance and post-annealing properties of coatings with different Nb contents were analyzed with the aim of developing RHEAS with excellent heat and wear resistance.

## 2. Experimental procedure

### 2.1 Theoretical calculations

In this work, first-principle calculations were used to determine various alloy properties (1)-(6) [25], including atomic radius difference (δ), entropy of mixing ($\Delta S_{mix}$), enthalpy of mixing ($\Delta H_{mix}$), thermodynamic parameter (Ω), valence electron concentration (VEC), and theoretical melting point of mixing ($T_m$). $\bar{r}$ represents the average atomic radius in the alloy, $C_i$ denotes the atomic fraction of the i$^{th}$ element in the alloy system, $r_i$ represents the atomic radius of the ith$^{th}$ element, R represents the gas constant (8.314 J/mol-K), $\Delta H_{ij}^{mix}$ represents the enthalpy of mixing between the two elements, $(VEC)_i$ denotes the valence electron concentration of the ith element, and (Tm)i corresponds to the melting point of the i$^{th}$ element.

$$\delta = \sqrt{\sum_{i=1}^{n} c_i \left(1 - \frac{r_i}{\bar{r}}\right)^2} \quad (1)$$

$$\Delta S_{mix} = -R \sum_{i=1}^{n} (c_i \ln c_i) \quad (2)$$

$$\Delta H_{mix} = 4 \sum_{\substack{i=1 \\ i \neq j}}^{n} \Delta H_{ij}^{mix} c_i c_j \quad (3)$$

$$\Omega = \frac{T_m \Delta S_{mix}}{|\Delta H_{mix}|} \quad (4)$$

$$VEC = \sum_{i=1}^{n} C_i (VEC)_i \quad (5)$$

$$T_m = \sum_{i=1}^{n} c_i (T_m)_i \quad (6)$$

The thermophysical parameters of RHEA with different Nb contents are shown in Table 2. According to the principle of solid solution formation, solid solution is formed at all Nb contents, which satisfies the conditions for HEAS formation. In the case of Nb$_x$Mo$_{0.5}$Ti$_{1.5}$Ta$_{0.2}$Cr, the variation of the relevant parameters with the value of x is shown in Fig. 1. It is clear that the most balanced theoretical properties are reached when the Nb content is 1.3 (at%), so x=1.3 is included in the experimental program.

**Table1 Atomic radius and mixing enthalpy between elements**

| ELement | Atomic radius /nm | Value of mixing enthalpy / (K·J·mol-1) | | | | |
|---|---|---|---|---|---|---|
| | | Nb | *Mo* | Ti | Ta | *Cr* |
| Nb | 0.146 | 0 | -6 | 2 | 0 | -7 |
| *Mo* | 0.139 | | 0 | -4 | -5 | 0 |
| Ti | 0.147 | | | 0 | 1 | -7 |
| Ta | 0.146 | | | | 0 | -24 |
| *Cr* | 0.128 | | | | | 0 |

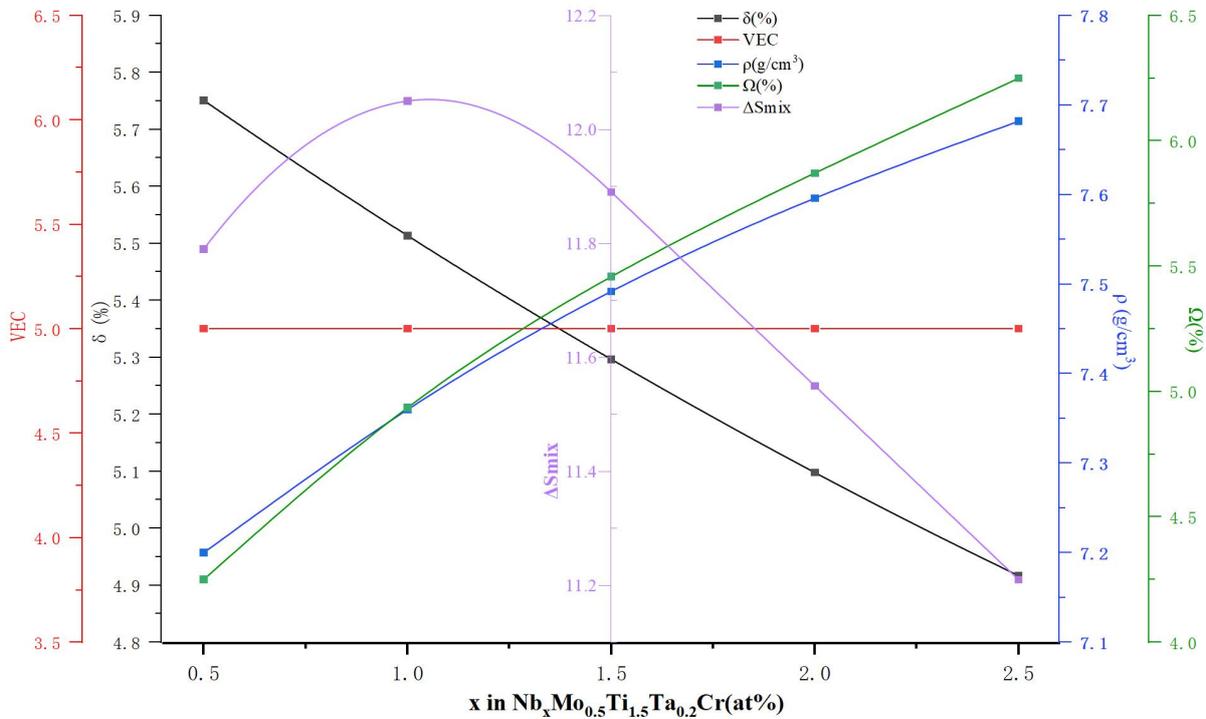

Fig. 1 Thermodynamic parameters change curve

**Table2 Thermodynamic parameters of alloy**

| Sample | $\delta_{(\%)}$ | △Smix (J/K mol) | △Hmix (KJ/mol) | $\Omega_{(\%)}$ | VEC |
|---|---|---|---|---|---|
| $Nb_{0.5}$ | 5.75 | 11.79 | -6.42 | 4.25 | 5 |
| $Nb_1$ | 5.51 | 12.05 | -5.78 | 4.93 | 5 |
| $Nb_{1.3}$ | 5.33 | 11.98 | -5.47 | 5.24 | 5 |
| $Nb_{1.5}$ | 5.29 | 11.89 | -5.25 | 5.45 | 5 |
| $Nb_2$ | 5.09 | 11.55 | -4.81 | 5.87 | 5 |

## 2.2 Preparation of the coating

The 316L high-temperature stainless steel was cut into 50 mm × 50 mm × 10 mm blocks, polished with 200- and 400-grade sandpaper, and cleaned with anhydrous ethanol. $Nb_xMo_{0.5}Ti_{1.5}Ta_{0.2}Cr$ (x = 1, 1.3, 1.5, 2) RHEAS powders with purity greater than 99.95% and particle size of 15-53μm were weighed and mixed, and then subjected to 10 h of vacuum ball milling. The powders (shown in Fig. 2) were mixed with 1 at% polyvinyl alcohol solution and compressed. The mixing $Nb_xMo_{0.5}Ti_{1.5}Ta_{0.2}Cr$ powders were preplaced on the substrate with the thickness of about 1 mm and then dried at 200 °C for 2 hours. An IPG YLS-3000 laser system was used to determine the optimal processing parameters after several trials using the controlled variable method: beam diameter of 3.6 mm, laser power of 2kw, scanning speed of 5 mm/s, offset of 3 mm, and the use of high-purity argon gas as a shielding layer.

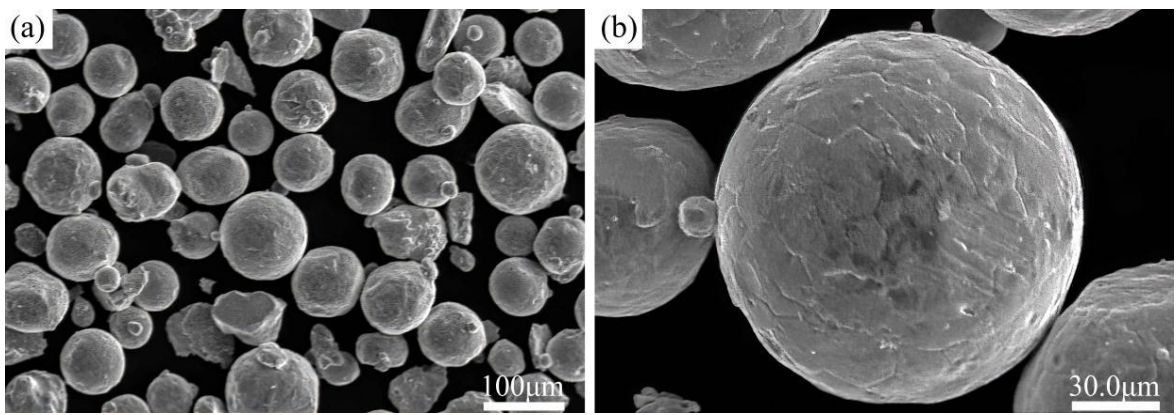

**Fig. 2 Microstructure of $NbMo_{0.5}Ti_{1.5}Ta_{0.2}Cr$ powder ( a ) Group morphology; ( b ) Single morphology**

## 2.3 Coating characterization

*Phase and Microstructure Analysis*

After etching with Keller's reagent (HF: HCl: HNO : $H_{32}$ O = 2: 3: 5: 100), 10 mm × 10 mm samples were prepared and analyzed using a SmartLab X-ray diffractometer and a JSM-7800F field emission scanning electron microscope. Elemental composition was evaluated using an equipped X-ray spectrometer.

*Mechanical and Wear Tests*

The surface microhardness was evaluated using a hmv-g21 Vickers hardness tester with a load of 1.961 N for 20 seconds, averaged over five measurements, each 100 μm apart. The coefficient of friction was measured using GCr15 tungsten carbide balls with a diameter of 4 mm, loaded with 5 N and rotated at 600 r-min for 25 min.

*Post-Annealing Analysis*

The samples were oxidatively annealed at 900°C for 10 hours and then repeatedly tested for hardness and friction properties to investigate high-temperature softening, strengthening mechanisms and oxidation conditions.

## 3. Results and Discussion

### 3.1 Crystal structures

Figure 3 shows the X-ray diffraction patterns of coatings with different Nb contents, there is one distinct peak and four secondary peaks in all Nb content coatings. The primary peak is located at 38.56° and the secondary peaks are located at 55.65°, 69.73°, 82.56° and 95.01°, which are roughly in the ratio of 1:2:3:4:5 with $\sin^2 \theta_1 : \sin \theta^2{}_2 : \sin \theta^2{}_3 : \sin \theta : \sin^2{}_4{}^2{}_5 \theta$, confirming the formation of the BCC phase [26]. Comparison with the JCPDS card D values indicates that these peaks are from the BCC phase formed by Nb-Ta rich clusters. Jade analysis shows a weak Mo-Nb peak structure to the right of the main peak of the BCC phase, with the main peaks located at 40.61°, 58.72°, 73.84°, and 87.70°, respectively, which is in agreement with the ratio standard of $\sin^2 \theta$ for the BCC2 phase.The BCC structure exhibits two different sets of peak phases in the figure, which can be attributed to the formation of disordered and chemically short-range ordered phases of the same group of elements at close intervals, a phenomenon that is due to inherent bias of elements. phenomenon is caused by the inherent polarization of the elements [27].

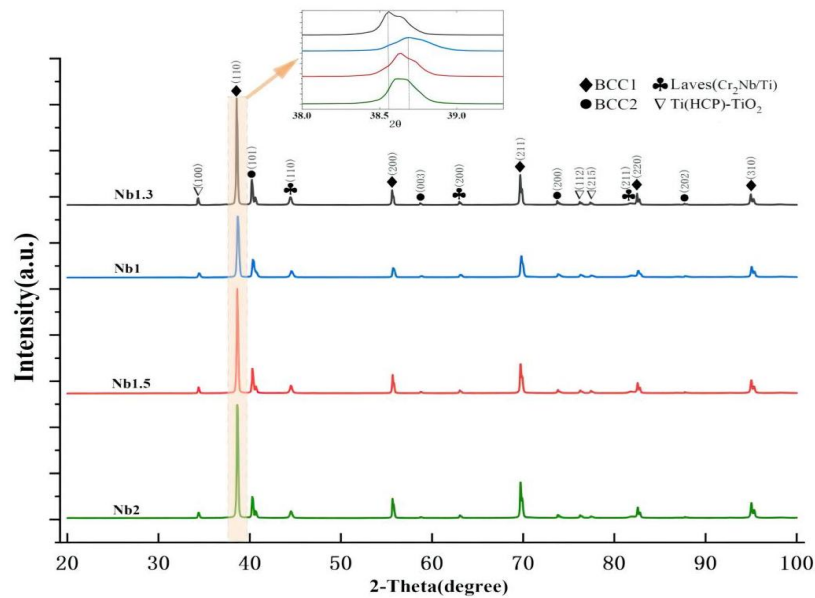

Fig. 3 XRD patterns of $Nb_xMo_{0.5}Ti_{1.5}Ta_{0.2}Cr$ RHEAS

Cr added by substitution has a small atomic radius and displaces other lattice elements, causing significant lattice distortions. Cr co-deposited in the BCC phase forms complex intermetallic compounds (Laves) with elements such as Nb and Ti , e.g. $C15(Nb,Ti)Cr_2$ . In addition, all coatings exhibit slight $Ti-TiO_2$ phase peaks.

HCP(Ti) is stabilized below 882°, while BCC(Ti) is stabilized above 882° [28]. At high laser transient temperatures, BCC(Ti) forms BCC phase with other BCC stable phase elements, but at the same time, the Ti element is less solid soluble, and during cooling, the HCP(Ti) element precipitates out to form the pure HCP (Ti) phase. During this transformation process, a small amount of airborne O elements penetrate into the coating, forming clusters of $TiO_2$ .

Figure 3 shows that the main peak shifts significantly to the left when the Nb content reaches 1.3. It is hypothesized that this is due to the Nb content reaching a critical value for this RHEAS lattice distortion. By entering the exact position of the main diffraction peak of the BCC phase and the corresponding crystallographic indices of the four coatings into Bragg's law ( $a=\dfrac{\lambda\sqrt{h^2+g^2+l^2}}{2\sin\theta}$ ) [29], we obtained the lattice constants "a" at the main peak for the four coatings, as shown in Table 3. At Nb1.3, the lattice distortion is significantly enhanced by the increase of lattice parameter and crystal Pitch. As shown in Fig. 4 [30], this phase transition involving volume changes may produce lattice dislocations and stress-induced cracking.

Table3 The main peak lattice constants at different Nb contents

| $Nb_x$ | $2\theta/°$ | (h k l) | $a_{(h k l)}$ |
|---|---|---|---|
| 1.3 | 38.562 | (1 1 0) | 3.2991 |
| 1 | 38.699 | (1 1 0) | 3.2878 |
| 1.5 | 38.640 | (1 1 0) | 3.2927 |
| 2 | 38.621 | (1 1 0) | 3.2942 |

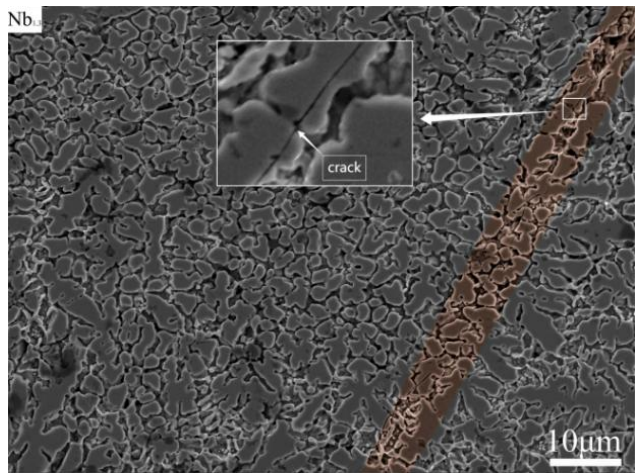
**Fig. 4 Crack at top of coating**

Figure 5 illustrates the macroscopic and microscopic morphology of RHEAS prepared by laser cladding at different coating heights. Fig. 5(a) shows a strong metallurgical bond between the coating and the substrate without visible defects such as cracks or pores. The microstructure of the top of the coating, as shown in Fig. 5(b), has a large number of "petal-like" grains, which is attributed to the compact granular structure caused by the temperature gradient and rapid cooling during laser cladding, which refines the grains and improves the mechanical properties. On the contrary, as shown in Fig. 5(c), the middle region of the coating exhibits an undirected equiaxed dendritic structure due to the loss of heat dissipation orientation. Fig. 5(d) shows a large number of coarse gray columnar dendrites with uniform orientation at the bottom of the coating, which is attributed to the accumulation of heat that makes the grains grow continuously and cannot be cooled in time to refine the grains. In addition, due to the dilution effect of the substrate, the saturation of the Cr element at the junction between the fusion cladding and the substrate promotes the grain growth.

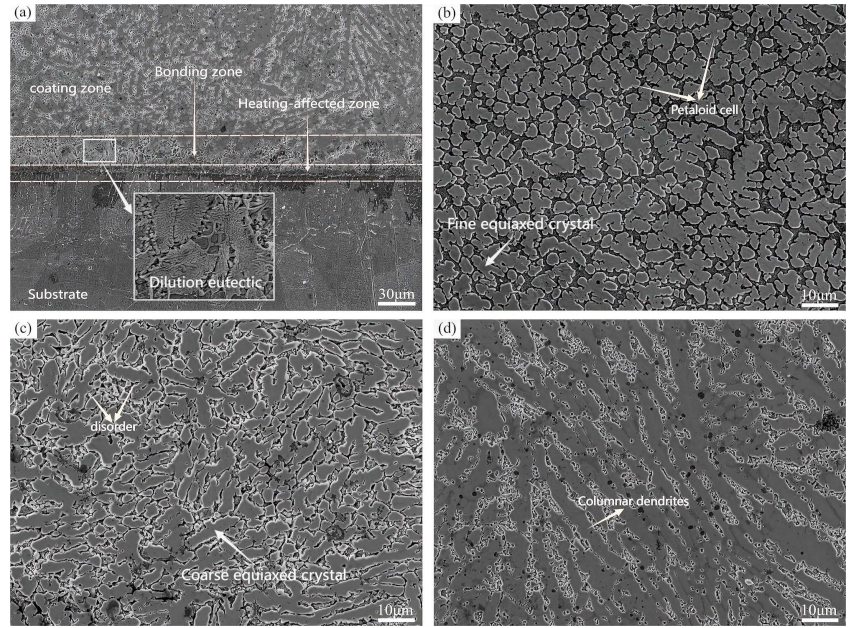

**Fig. 5 Macroscopic view of alloy coating bonding zone and microstructure of different cladding layer heights (a) Bonding zone (b) Top of cladding layer (c) Middle of cladding layer (d) Bottom of cladding layer**

Fig.6 shows the microstructure of the middle part of the Nb$_x$Mo0.5Ti1.5Ta0.2CrRHEAS coating. As shown in Fig. 6(b-e) the coatings are mainly composed of dendrites with equiaxed crystals. With the increase of Nb content, the number of rod-like dendrites gradually occupies the interstitial region of the matrix. The microstructure of alternating growth of light gray phase, white dendritic phase, and dark gray phase appears in Nb$_{1.3}$ shown in Fig. 6(a). The high solidification point of Ta induces the first nucleation of the BCC1 phase, which forms "snowflakes" or equiaxed dendrites similar to "snowflakes" or equiaxed dendrites, which is in agreement with the light gray portion in the image. During this process, the entire coating remains in a solid-liquid mixed state. The not-yet-solidified Mo-Nb-rich phase surrounds the periphery of the BCC1 phase as the temperature decreases, or aggregates in the region between the dendrite arms to form the bright white BCC2 phase. On the other hand, Cr, Nb and Ti, which have relatively low melting points, precipitate out of the liquid phase and fill up between the BCC phases to form the Laves layer. As shown in Fig. 6(e), an increase in Nb content increases the prevalence of the light gray BCC1 phase while decreasing the bright BCC2 phase. In addition, it is observed that the increase in Nb content also leads to significant grain refinement, which is attributed to the effect of diffuse plasmas promoting non-homogeneous nucleation. This verifies the inference of Table 3 and Fig. 3.

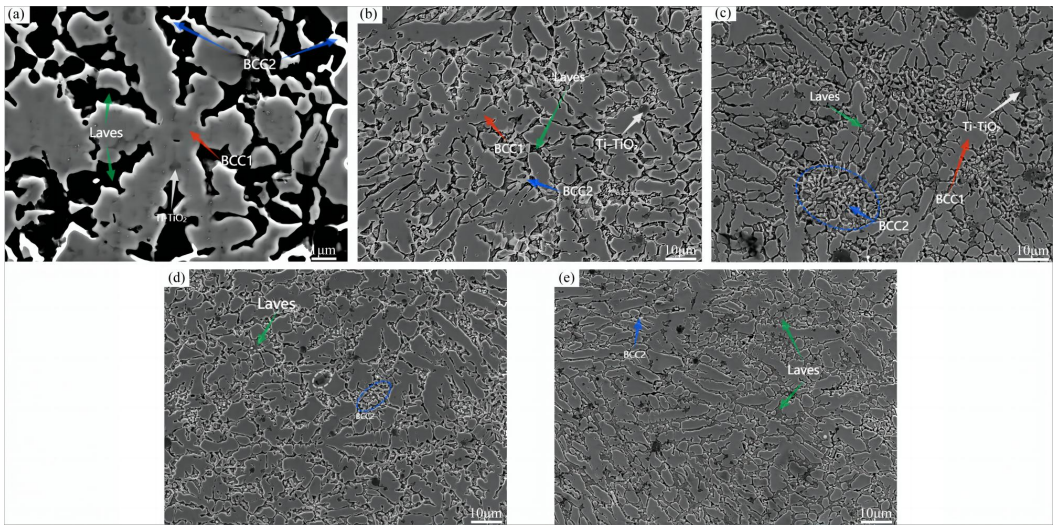

**Fig. 6 SEM images of $Nb_xMo_{0.5}Ti_{1.5}Ta_{0.2}Cr(x=1.3,1,1.5,2)$RHEAS(a-b)$Nb_{1.3}$(c)$Nb_1$(d)$Nb_{1.5}$(e)$Nb_2$**

Fig.7 Elemental distributions in the microstructure of $Nb_xMo_{0.5}Ti_{1.5}Ta_{0.2}Cr$ were analyzed using energy dispersive spectroscopy (EDS). As shown in Fig. 7(b)-(d), the dendrites are mainly composed of Nb, Ta and Mo, and the distribution of Ta is obviously narrower, mainly in the light gray part of the SEM image. The appearance around the edge of the dendrite represents the continuous BCC2 phase, which is mainly formed by Nb-Mo. This bright BCC2 phase also penetrates into the interstices of the main dendrite, reflecting the transition of the BCC1 phase grain boundaries due to the solid-state phase transition during solidification. As shown in Fig. 7(f), the high-temperature BCC biphasic phase inhibits Cr from entering grain boundaries,

precipitating, and concentrating it in the dendritic interstices throughout the cooling process. As shown in Fig. 7(e), the Ti that can form a biphasic phase at different temperatures is uniformly distributed throughout the coating, and the bright bulk portion indicates the accumulation of oxides during the crystallization process.

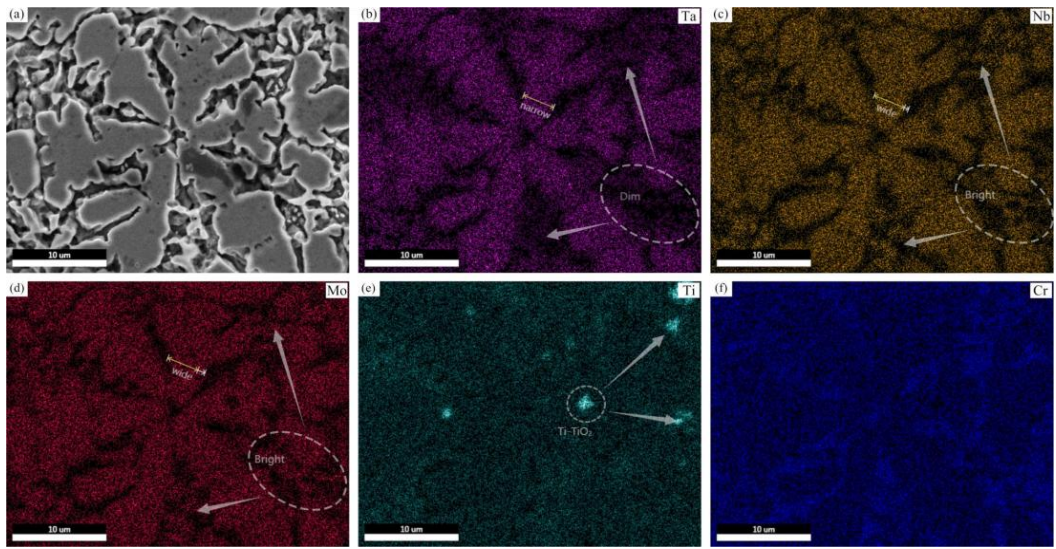

Fig.7 SEM images and EDS mapping of component for $Nb_xMo_{0.5}Ti_{1.5}Ta_{0.2}Cr$ RHEAS(b)Ta(c)Nb(d)Mo(e)Ti(f)Cr

In order to assess the high temperature stability of this RHEAS structure and organization, Fig. 8 shows the SEM morphology of the coatings with different Nb contents after oxidative heat treatment at 900°C for 10 hours. The results show that the grain morphology after annealing at 900°C remains essentially the same as before annealing, maintaining the equiaxed grains. However, there is a shift from thicker equiaxed dendrites to thinner and shorter rods. This is attributed to the formation of highly stable clusters by the Nb-Ta-Mo elements [31], which prevented any significant phase transitions within the lattice, thus ensuring the high-temperature stability and high-temperature creep resistance of the coating. A significant reduction in the bright BCC2 phase was observed compared to the pre-annealing phase, which can be attributed to the gradual penetration of Ta elements from the BCC1 phase into the BCC2 phase. The Mo element, which was initially homogeneously distributed in the double BCC phase, remains relatively stable due to its inherent excellent creep resistance.

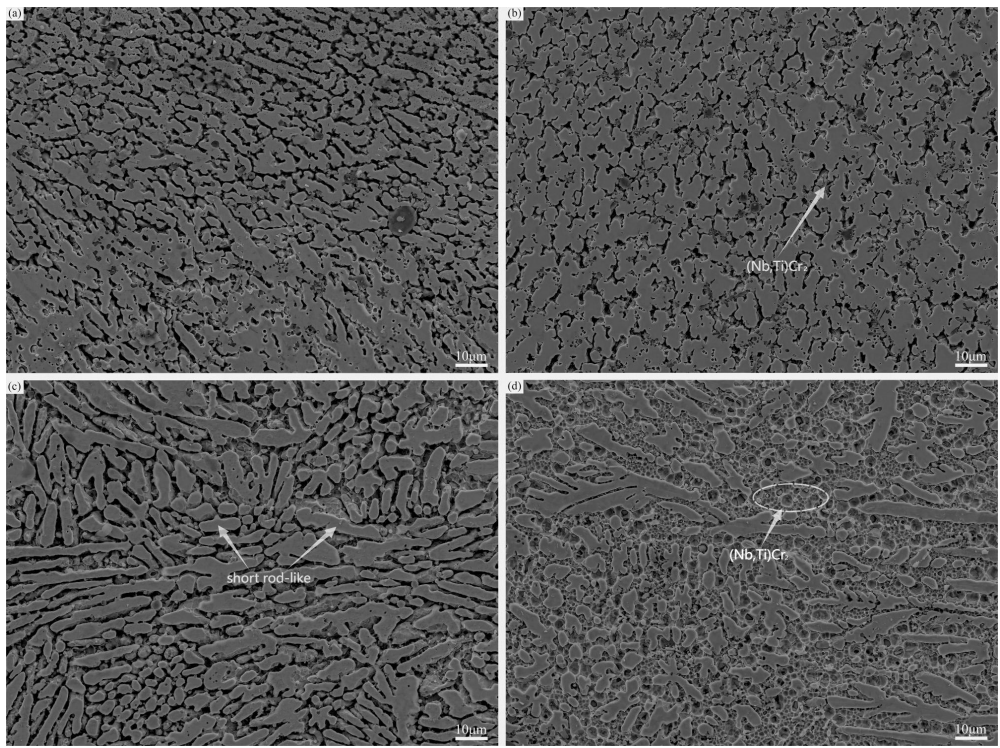

**Fig. 8 SEM images of $Nb_xMo_{0.5}Ti_{1.5}Ta_{0.2}Cr(x=1.3,1,1.5,2)$RHEAS after annealing (a)$Nb_{1.3}$(b)$Nb_1$(c)$Nb_{1.5}$(d)$Nb_2$**

With increasing Nb content, Fig.8(b)-(d) shows an increasing presence of (Nb,Ti)$Cr_2$ Laves phase. This phenomenon originates from the enrichment of Nb elements in the interstitial region of the matrix, which nucleate with Cr elements at high temperatures, thus forming the Laves phase. The EDS analysis after annealing is shown in Fig.9 , where it is observed that a significant amount of Ti precipitates into the inter-dendritic matrix portion. This is due to the fact that the annealing temperature is 900 °C, which is close to the phase transition temperature of Ti at 882 °C. As a result, the Ti element is in the critical state of phase transition for a long period of time, which reduces the strength of the interatomic metallic bonds. Upon cooling, the Ti element finally precipitates [28] and subsequently interacts with the excess Cr in the x=2 alloy to form the $TiCr_2$ Laves phase.

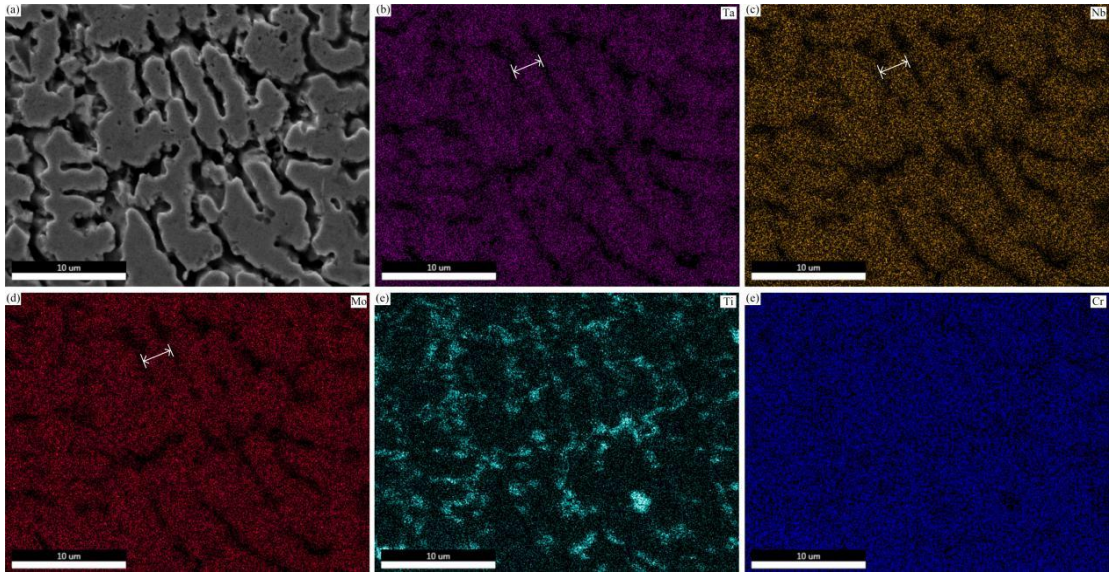

**Fig. 9 EDS mapping of component for $Nb_xMo_{0.5}Ti_{1.5}Ta_{0.2}Cr$(x=1.3,1,1.5,2)RHEAS after annealing (a)$Nb_{1.3}$(b)$Nb_1$(c)$Nb_{1.5}$(d)$Nb_2$**

### 3.2 Coating microhardness analysis

The variation of $Nb_xMo_{0.5}Ti_{1.5}Ta_{0.2}Cr$ coating hardness with height is shown in Fig. 10(a). A significant increase in microhardness was observed compared to the substrate, with the most pronounced increase at an Nb content of 1.3. It is hypothesized that solid solution strengthening is most pronounced at this concentration. According to Table 3, at this time, the lattice parameter become larger, leading to an increase in the crystal face spacing, which in turn induces lattice dislocations. Lattice dislocations are linear defects that are usually undesirable in processes such as vacuum melting. However, understanding and controlling these dislocations in laser cladding processes can improve the mechanical properties of coatings [32]. However, it must be noted that dislocations may induce defects such as cracks on the coating surface.

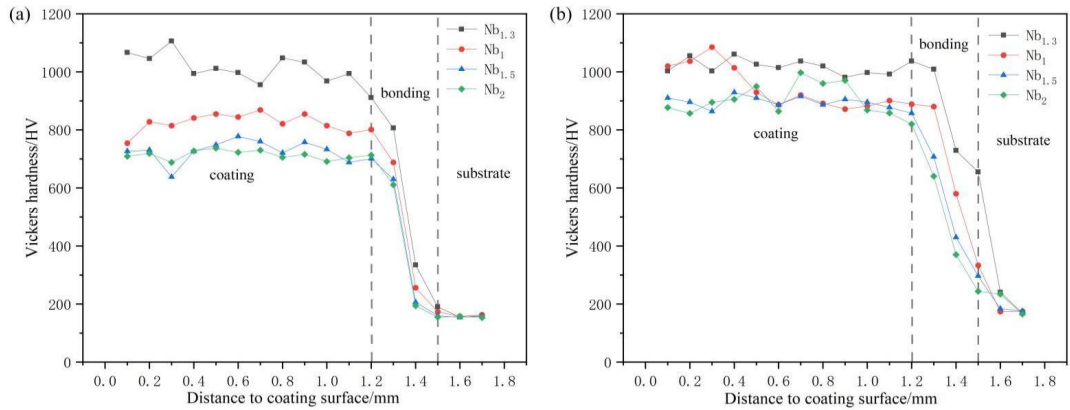

Fig. 10 Microhardness of laser cladding layer: ( a ) before annealing; ( b ) after annealed

For other Nb concentrations, with the increase of Nb content, the concentration of Cr and Mo elements in the alloy solution decreases, the solid solution strengthening effect is gradually weakened, and the microhardness of the coatings also decreases gradually. Before annealing, when $Nb_{1.3}$, the peak hardness reaches 1066.5HV, which is about 7.11 times of the hardness of the base material. Even at $Nb_2$, the peak hardness reaches 713.8 HV, which is about 4.75 times the hardness of the base material. This is mainly attributed to the inherent hardness of the BCC phase and the relatively high hardness of the $Cr_2$ (Ti/Nb) Laves phase as well. The coatings microhardness increased significantly after annealing due to the refinement of grain. When $Nb_2$, the enhancement is as high as 31.2%. This can be attributed to the excess Nb leading to a large generation of Laves phase, as shown in Fig. 8(d), where significant grain refinement is accompanied by an increase of Laves phase in the dendrites. Therefore, it is clear that this RHEAS has excellent resistance to high-temperature softening.

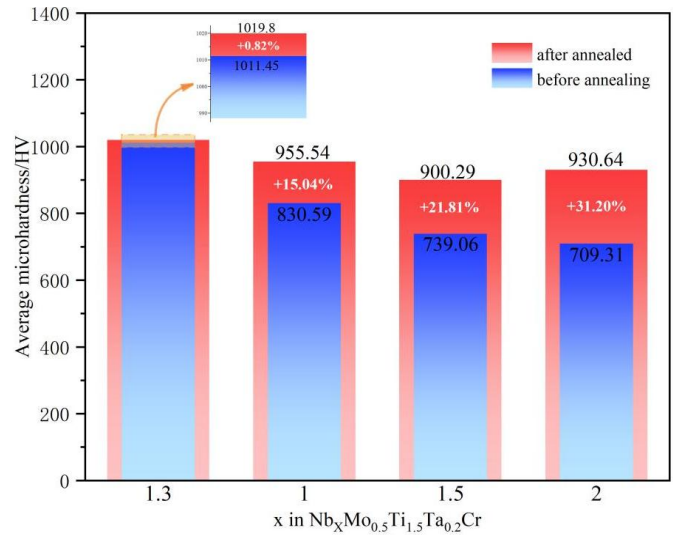

Fig. 11 Average hardness of coating before and after annealing

### 3.3 wear resistance

Figure 12 depicts the coefficient of friction of 316L stainless steel matrix and $Nb_xMo_{0.5}Ti_{1.5}Ta_{0.2}Cr$ RHEAS at different Nb contents. Fig. 12(a) shows the coefficient of friction before annealing, which shows on the surface that the first 6 minutes are an unstable wear phase. After that, a smoother fluctuating state develops due to localized fracture of the worn surface and cyclic accumulation and expulsion of wear debris.

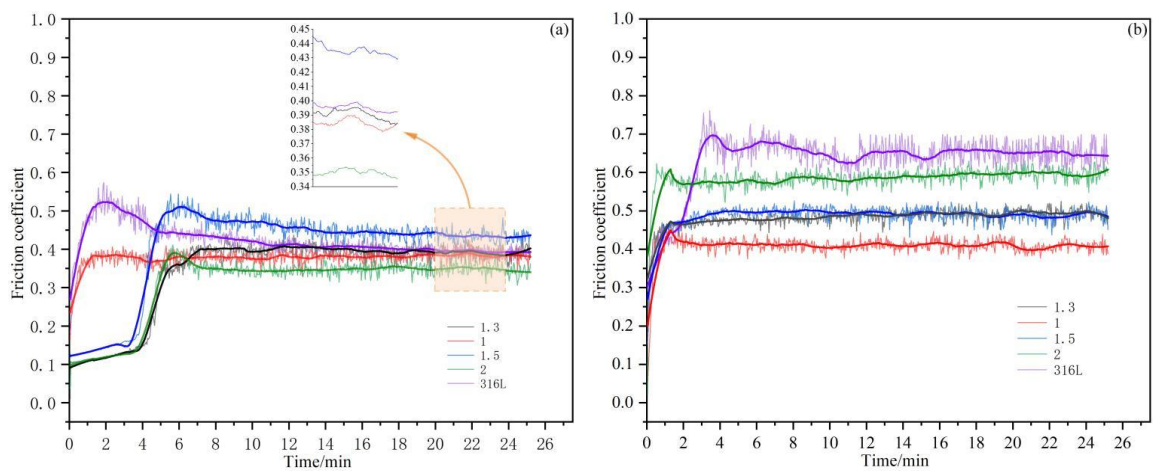

Fig. 12 The friction and wear curves of 316L stainless steel and RHEAS coating. ( a ) before annealing ( b ) after annealed

Before annealing, the coefficient of friction of the $Nb_{1.5}$ coating is higher than that of the base material 316L stainless steel in the steady state phase. The coefficients of friction of the $Nb_1$ and $Nb_{1.3}$ coatings are slightly lower than that of the base material. The coefficient of friction of the $Nb_2$ coating is considerably lower, which indicates minimal surface wear and plastic deformation. Measurement of the width of the wear marks by SEM on the wear mark wheels reveals that the differences between the different Nb contents are minimal and there is no clear pattern. Among them, the wear tracks of $Nb_{1.5}$ and $Nb_2$ are wider, which may be due to their relatively lower hardness, resulting in wider wear track widths. However, in Fig. 13(a), the RHEAS with the lowest hardness (x=2) shows the lowest average friction coefficient. According to classical wear theory [33], an increase in hardness significantly improves wear resistance, which is counterintuitive.

The coefficient of friction of $Nb_xMo_{0.5}Ti_{1.5}Ta_{0.2}Cr$ after annealing is shown in Fig. 12(b). After annealing, the friction coefficient of 316L increased by 66.34%. The coating friction coefficients at all Nb contents are lower than the friction coefficients of the base material. It is noteworthy that the coefficient of friction is the smallest when $Nb_1$, which is only increased by 7.63% compared with that before annealing, which indicates excellent high temperature friction and wear performance. On the contrary, when $Nb_2$, the friction performance after annealing decreases drastically and the coefficient of friction increases relatively by 68.19%. As can be seen from Fig. 13(b), the wear of 316L stainless steel substrate and $Nb_xMo_{0.5}Ti_{1.5}Ta_{0.2}Cr$ RHEAS (x-values of 1.3, 1, 1.5, and 2, respectively) are 4.87, 1.36, 0.76, 1.44, and 3.92, respectively. It is clear that the alloy coating shows excellent wear resistance with much less wear compared to the substrate.

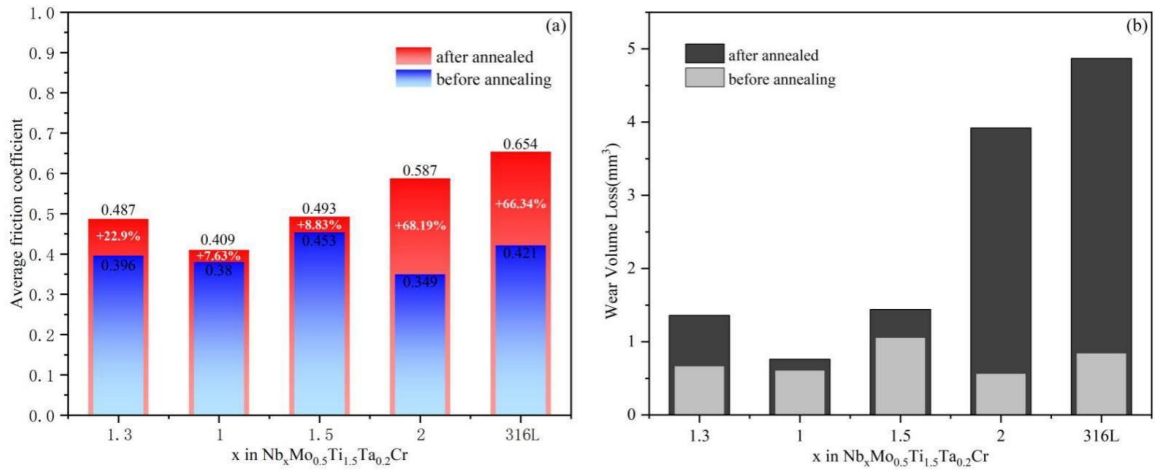

Fig. 13 Average friction coefficient and wear mass loss (a) Average friction coefficient (b) wear mass loss

As shown in Fig. 13(a), the frictional wear pattern of 316L shows many wide and deep grooves in the substrate, along with obvious flaking pits and debris marks, darker bonded areas, and significant plastic deformation, all of which indicate severe wear.EDS measurements show elevated oxygen levels in the darker flake areas, suggesting that oxidative wear is present.316L stainless steel is significantly harder than GCr15 (~800HV), and the low hardness makes it susceptible to plastic V deformation when in contact with harder materials, leading to cracking. GCr15 (~800HV), and its low hardness makes it susceptible to plastic deformation in contact with harder materials, which can lead to cracking. As the friction process continues, the temperature of the coating surface rises, eventually forming an oxide layer that subsequently peels away from the substrate. Therefore, the predominant wear pattern of 316L prior to annealing appears to be adhesive wear, supplemented by oxidative wear.

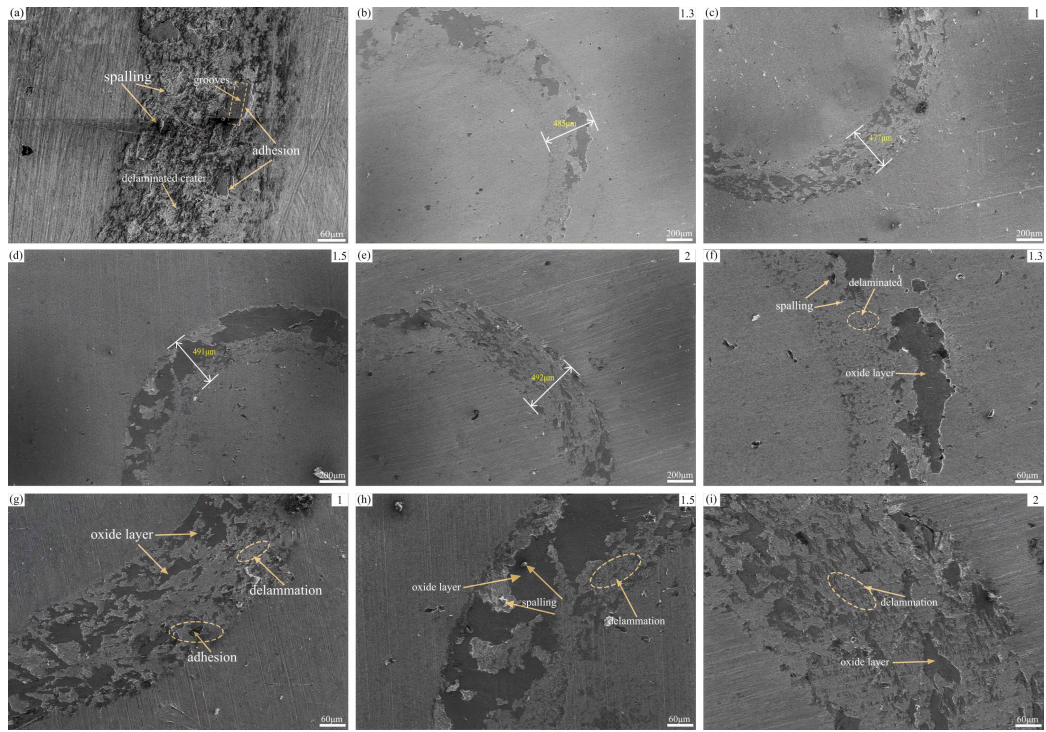

**Fig. 14 Morphology of wear trajectory. (a) matrix; (b) ~ (e) before annealing x=1.3,1,1.5,2; (f) ~ (i) after annealing x=1.3,1,1.5,2**

Observation of the wear morphology in Figs.14(f) to 14(i) reveals that due to the significant increase in coating hardness, the grooves in the wear morphology become significantly lighter, while the darker layer areas become more prominent. This continuous and tightly packed oxide layer formed during friction exhibits only subtle groove features. The oxide layer is generated in situ on the alloy surface during pre-annealing friction and acts as a barrier between the coating and the GCr15 balls, inhibiting adhesion and enhancing wear resistance. Therefore, the predominant wear pattern before annealing is oxidative wear accompanied by a small amount of abrasive wear. It is noteworthy that the highest hardness but not better friction properties were obtained at $Nb_{1.3}$, which is not consistent with the classical wear theory. The reason for this inconsistency is that as the hardness reaches a certain critical value, the alloy becomes more brittle, leading to cracking and resulting in severe abrasive grain exploitation, which in turn leads to severe abrasive wear. In contrast, the coating at $Nb_2$ exhibits superior toughness, reducing the likelihood of debris and crack formation. Observation of the wear trajectories reveals that the oxide layer at $Nb_{1.3}$ and $Nb_{1.5}$ is very heterogeneous. This oxide layer is formed by debris that builds up and compresses during the friction process. Once these buildups are formed, they are prone to internal cracking and show undesirable wear resistance. In contrast, at $Nb_1$ and $Nb_2$, the oxides generated in situ in the sliding direction are point-like and densely and uniformly distributed, which greatly improves the wear resistance.

Upon annealing, the substrate surface precipitates large amounts of iron, Cr and nickel carbides in the austenite, leading to a drastic reduction in wear performance. On the contrary, the oxide film on the RHEAS surface is mainly composed of various oxides such as $Nb_2O_5$, $TiO_2$ and $Ta_2O_5$. According to the X-ray diffraction pattern of the oxide layer shown in Fig. 15(a), the highest percentage content of Ti and Ta in the coating was observed when $Nb_1$ was present. According to the peak intensity comparison, the coating formed a denser oxide film at $Nb_1$ than at $Nb_2$. Among them, $Ta_2O_5$ has relatively high hardness (about 10 GPA) and can be used as a solid lubricant due to its orthorhombic structure, which makes it easy to form shear surfaces. In addition, the dense $TiO_2$ oxide film also acts as a lubricant and effectively reduces the coefficient of friction (COF). The synergistic effect of the Ta and Ti oxides significantly improves the wear resistance of the alloy. However, one of the reasons for the increase in friction coefficient of the $Nb_2$ coating, which had the lowest COF before annealing, is the significant increase in hardness after annealing (31.2%), which leads to the formation of cracks in the surface layer of the coating. In addition, too high Nb content leads to the formation of complex intermetallic oxides of $Ti_4Cr_3NbO_{32}$ type, which is detrimental to the wear resistance.

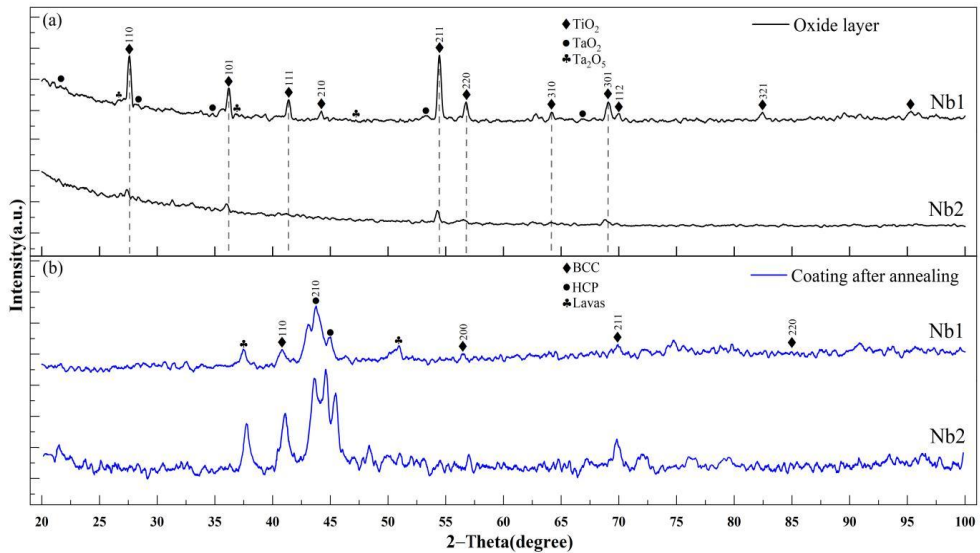

Fig.15 XRD pattern after annealing ( a ) Oxide layer ( b ) Coating

## 4. Conclusions

Utilizing Laser Cladding (LC) technology, refractory high-entropy alloy coatings (RHEAcs) of $Nb_xMo_{0.5}Ti_{1.5}Ta_{0.2}Cr$ were successfully fabricated on the surface of 316L stainless steel. Through systematic studies, the phase compositions, structural characteristics, and the impact of Nb content on hardness and wear resistance were elucidated. This coating boasts exceptional mechanical properties, lower density, and production costs, heralding a promising application prospect. The principal conclusions are as follows:

(1) The composition of RHEAcs primarily consists of a body-centered cubic (BCC) solid solution, C15-Laves phase, and a minor quantity of hexagonal close-packed (HCP) solid solution containing Ti. As the Nb content increases, the crystal grains exhibit a progressively refined trend.

(2) In comparison to the base material, the RHEAcs coating demonstrates a significant enhancement in hardness. When $Nb_{1.3}$, influenced by lattice dislocations, the coating's hardness ascends to 1066.5HV, which is 7.11 times that of the base material. However, at $Nb_2$, despite the hardness being the lowest at 709.31HV, it still reaches 4.72 times that of the base material. After annealing, the hardness across different Nb contents all exhibited improvement, particularly at $Nb_2$, where the hardness increased by 31.2%.

(3) Before annealing, the wear performance of the RHEAcs coating is generally slightly superior to that of 316L stainless steel. However, post-annealing, the coating's coefficient of friction (COF) and wear rate are significantly lower than that of annealed 316L stainless steel, reflecting its superior wear resistance. Specifically, the coating at $Nb_1$, post-annealing, displays optimal wear resistance with minimal difference compared to before annealing. Furthermore, at $Nb_1$, the theoretical density of the coating is the lowest, standing at 7.3g/cm3. The outstanding wear resistance after annealing can be attributed to the ultra-fine grain oxidation friction layer formed, such as $TiO_2$ and $Ta_2O_5$. The primary wear mechanism involves oxidative wear, accompanied by a minor amount of abrasive wear.

## Acknowledgments

The authors greatly appreciate the financial support from the National Natural Science Foundation of China (Nos. 51875267), Young Key Teachers Program of Jiangsu University.